\documentclass[preprint,aps,showpacs]{revtex4}
\usepackage{epsfig}
\usepackage{graphicx}
\usepackage{bm}
\usepackage{amssymb}
\usepackage{dcolumn}
\usepackage{bm}
\begin{document}
\draft

\title
{Magic polarization for optical trapping of atoms without Stark-induced dephasing}
\author{Huidong Kim, Hyok Sang Han, and D. Cho\footnote{e-mail address:{\tt cho@korea.ac.kr}}}
\affiliation{Department of Physics, Korea University, Seoul 136-713, Korea}

\date{\today}

\begin{abstract}
We demonstrate that the differential ac-Stark shift of a ground-state hyperfine transition in an optical trap can be eliminated by using properly polarized trapping light. We use the vector polarizability of an alkali-metal atom to produce a polarization-dependent ac-Stark shift that resembles a Zeeman shift. We study a transition from the $|2S_{1/2}, F=2, m_F =-2 \rangle$ to the $|2S_{1/2}, F=1, m_F  =-1 \rangle$ state of $^7$Li to observe $0.59 \pm 0.02$ Hz linewidth with interrogation time of 2 s and $0.82 \pm 0.06$ s coherence time of a superposition state. Implications of the narrow linewidth and the long coherence time for precision spectroscopy and quantum information processing using atoms in an optical lattice are discussed.
\end{abstract}


\maketitle

An optical trap \cite{Chu PRL 1986} is the least perturbative among atom traps, providing promising environment for precision spectroscopy. An optical lattice, formed by interfering light fields, is an attractive platform for quantum information processing (QIP) for the possibility of site-specific qubit manipulation and scalability \cite{Bloch Nature 2008}. Utility of these devices is severely limited, however, due to a differential ac-Stark shift coupled with an intensity gradient of the trapping field. For an optical transition, the ``magic wavelength" eliminated the differential shift \cite{Cho JKPS 2003, Ye Science Review} leading to breakthroughs in optical frequency metrology \cite{Katori PRL 2003} and cavity quantum electrodynamics \cite{Kimble PRL 2003}. In contrast, for a ground hyperfine transition, which is used as the primary frequency standard and a resource for a qubit, the problem remained unsolved.

Considering two ground hyperfine states $|\phi_l \rangle = |nS_{1/2}, F \rangle$ and $|\phi_u \rangle =|nS_{1/2}, F+1 \rangle$ of an alkali-metal atom with $F$ and $F+1$ being total angular momenta, the ac-Stark shift produces different potential wells for the two states because of the hyperfine splitting $\Delta_{\rm HF}$. A trapped atom has an eigenstate $|\phi_l \rangle |\chi_{ln} \rangle$ or $|\phi_u \rangle |\chi_{um} \rangle$, where $|\chi_{ln} \rangle$ and $|\chi_{um} \rangle$ are the motional bound states. Because the potential wells are different, the bound states have different sets of energy eigenvalues and the orthogonality $\langle \chi_{ln} | \chi_{um} \rangle = \delta_{nm}$ is not satisfied. For spectroscopy on an ensemble of atoms, this gives rise to a resonance frequency shift and inhomogeneous broadening \cite{Cho JKPS 2007}, nullifying the advantage of long interrogation. For a superposition state, spatial and thermal distributions of resonance frequencies lead to fast dephasing \cite{Meschede PRA 2005}. Even for a single atom, parasitic excitations to adjacent motional states can degrade coherence in a short time. The dephasing has been the main roadblock in using a pair of hyperfine states as a qubit in an optical-lattice based QIP; an entangled state formed by cold collisions lost coherence in 1.5 ms \cite{Bloch Nature 2003} and the number of steps in a quantum-walk experiment was limited to 6 \cite{quantum walk}.

There has been a long series of efforts to address this problem. A blue-detuned optical bottle was used to minimize atom-light interaction \cite{Chu PRL 1995}. A near-resonant compensating beam was added to reduce the differential shift \cite{Davidson PRA 2002, Kuzmich Nature Physics 2010}. These arrangements cannot form a lattice and the near-resonant compensating beam increased scattering of photons. Spin echo or its generalization, dynamical decoupling, was used to undo Stark-induced dephasing \cite{Davidson PRL 2010}. Although it produced long storage time of hyperfine coherence for a quantum memory, the process is disruptive to spectroscopic measurements and gate operations in QIP. Ultracold atoms from a Bose-Einstein condensate were loaded to the ground motional states of a shallow optical lattice  \cite{Bloch PRL 2009}. Although this circumstance minimizes the dephasing due to motional distribution, problems from parasitic excitations and spatial distribution of atoms remain. An ideal solution would be a simple knob to adjust an optical trap so that the pair of hyperfine states have identical potential wells, as tuning a wavelength does for an optical transition \cite{Cho JKPS 2003}. Polarization of the trap beam can be such a knob \cite{Cho JPCS 2007} and in this Letter we show that by tuning polarization of the trap beam we can eliminate the differential Stark shift at its origin. We may call it ``magic polarization".

When an atom in the $|nS_{1/2}, F, m_F \rangle$ state, with $m_F$ being a $z$-component angular momentum, is in a laser field propagating along the $z$-axis with an amplitude $\vec\mathcal{E}$, its ac-Stark shift is
\begin{equation}
U({F, m_F}) = \alpha_F |\vec\mathcal{E}|^2 + \beta_F \eta g_F m_F |\vec\mathcal{E}|^2 ,
\end{equation}
where $\alpha_F$ and $\beta_F$ are the scalar and vector polarizabilities, respectively. The second term is from the spin-orbit coupling and it resembles a Zeeman shift; $g_F$ is the Land{\'{e}} $g$ factor and $B_{e\!f\!f}=\beta_F \eta  |\vec\mathcal{E}|^2/\mu_B$ plays the role of an effective magnetic field. $\eta=i \hat{z}\cdot (\vec\mathcal{E} \times \vec\mathcal{E}^*)/|\vec\mathcal{E}|^2$ is the degree of circularity and $\mu_B$ is the Bohr magneton. For a transition from the $|F, m_F \rangle$ to the $|F+1, m_F^\prime \rangle$ state, the differential Stark shift vanishes when
\begin{equation}
\eta = \frac{\Delta \alpha}{\beta g_F (m_F+m_F^\prime)},
\end{equation}
where $\Delta \alpha = \alpha_{F+1} - \alpha_F$ and $\beta = (\beta_{F+1}+\beta_F)/2$.  When light propagation makes an angle $\theta$ to the $z$-axis defined by a magnetic field, $\eta$ depends on $\cos \theta$ and the angle can be another knob \cite{Derevianko PRL 2008}. Magic polarization  works only for a Zeeman-sensitive transition. It is a liability in a fluctuating magnetic field, but it is an asset for addressing or manipulating qubits using a magnetic field gradient \cite{Meschede NJP}. For a Zeeman-insensitive transition with $m_F^\prime = -m_F$, the second-order Zeeman shift contributed by $B_{e\!f\!f}$ can cancel the differential shift \cite{Derevianko PRL 2010}. The scheme was applied to a two-photon transition of $^{87}$Rb with $m_F =1$ for a metrological application \cite{Porto PRL 2011}.

From Eq. (1), if we require that the difference in the potential well depths for the $|2S_{1/2}, F, m_F \rangle$ and  $|2S_{1/2}, F+1, m_F^\prime \rangle$ states, $\Delta U = U({F+1, m_F^\prime})- U({F,m_F})$, be smaller than $\delta U$, deviation $\delta \eta$ of the circularity from the optimal value in Eq. (2) should satisfy $\delta \eta \leq (\delta U/\bar{U})(\alpha/\beta g_F (m_F+m_F^\prime))$, where $\bar{U}$ is the average well depth. When the strengths of the $D1$ and $D2$ couplings are the same, $\alpha/\beta \approx 3\Delta/\Delta_{\rm F}$, where $\Delta$ is the detuning of the trap beam from the $D$ transitions and $\Delta_{\rm F}$ is the fine structure ($\Delta \gg \Delta_{\rm F}$); tolerance on $\eta$ in tuning out the differential shift is inversely proportional to the fine structure.  We use $^7$Li for its small fine structure of 10 GHz.

A double magneto-optical trap (MOT) system fed by a Zeeman slower \cite{ZSL} is used to load an optical trap. The optical trap is formed at the center of a glass chamber by focusing a 6-W single-frequency Gaussian beam with a wavelength $\lambda$ of 1060 nm to a spot of 20-$\mu$m intensity radius. The well depth is 600$E_R$ (0.7 mK) with $E_R =h^2/2m_{\rm Li}\lambda^2$, and $m_{\rm Li}$ is $^7$Li mass. $1.4 \times 10^{4}$ atoms at temperature $T$ of 140 $\mu$K are trapped in 1 s. Polarization of the trap beam is controlled by a quarter waveplate. When the waveplate optical axis is rotated by $\theta_{\rm QWP}$ with respect to the trap beam polarization direction, $\eta = \sin(2\theta_{\rm QWP})$. We use a motorized rotation stage (Thorlabs Model PRM1Z8) with 0.1$^\circ$ repeatability, which allows us to set $\eta$ to within $3.2 \times 10^{-3}$. Pressure inside the chamber is $8 \times 10^{-11}$ mbar and the trap lifetime for atoms in the $F=1$ and $|F=2, m_F =\pm 2 \rangle$ states is 20 s. For atoms in the $|F=2, m_F =0, \pm 1 \rangle$ states, however, it is 5 s owing to the hyperfine exchange collisions. The trapped atoms are optically pumped to the $|2, -2 \rangle$ state using the $D1$ transitions. The radio-frequency (RF) field at $\Delta_{\rm HF}$ = 803 MHz from a frequency synthesizer, which is phase locked to a rubidium atomic clock, is applied to a patch antenna placed under the glass chamber to drive a transition to the  $|1, -1 \rangle$ state. A 16-mG magnetic field along the trap beam defines the quantization axis.  The hyperfine transition cannot be detected by using a cycling transition because of the small $2P_{3/2}$ hyperfine splitting of lithium. Instead, we use a magnetic trap for the detection; atoms in the  $|1, -1 \rangle$ state are held by the trap, whereas those in the  $|2,-2\rangle$ state are pushed away. After holding the atoms in the magnetic trap with the optical trap turned off for 130 ms, the MOT beams are turned back on. For the magnetic trap, we use the anti-Helmholtz coil for the MOT with the same field gradient of 26 G/cm along the coil axis. The detection efficiency is close to 1. Fluorescence from the MOT is collected and collimated by an aspheric lens, and it is refocused to a photodiode via a pin hole. In order to improve the signal-to-noise ratio, we actively stabilize the MOT trap power and modulate the repump frequency by 16 MHz at 15 Hz. By integrating the demodulated fluorescence signal for 1 s, we can detect 200 atoms within one standard deviation.

Besides the ac-Stark shift, photon scattering and magnetic field variation can lead to broadening and decoherence. Raman scattering is suppressed from a destructive interference between the $D1$ and $D2$ paths \cite{Raman}, and the Raman decoherence rate is negligibly small for our far-detuned trap. While the elastic scattering rate is 3 s$^{-1}$, most of the scatters are benign. The Rayleigh decoherence rate is proportional to $1/\Delta^4$ \cite{Rayleigh} and is negligible as well. Although the transition is Zeeman sensitive, the malignant effect of a magnetic field gradient is reduced for the trapped atoms in a manner similar to the Lamb-Dicke effect. In a harmonic approximation, a gradient $bz\hat{z}$ translates and shifts a potential well without changing its shape; transition frequencies from the $|\phi_l \rangle |\chi_{ln} \rangle$ to $|\phi_u \rangle |\chi_{um} \rangle$ state with $m=n$ shift uniformly. A transition amplitude for $m=n\pm1$ is reduced by $\xi \sqrt{n}$ compared with the $m=n$ transition, where $\xi= \Delta z(0)/ \Delta z_b$; $\Delta z(0)$ is the size of the ground wave packet, and $\Delta z_b$ is the displacement at which differential Zeeman shift between $|\phi_l \rangle$ and $|\phi_u \rangle$ is the same as the motional excitation energy $\hbar \omega_z$. $\xi$ is analogous to the Lamb-Dicke parameter $k\Delta z (0)$, where $k$ is a wavenumber for an interacting photon. We minimize the effect of an ambient field by placing the glass chamber with accompanying optics and coils inside a 3-layer magnetic shield. We measure $b$ by mapping the resonance frequency while the optical trap is translated along the $z$-axis and keep it smaller than 0.25 $\mu$G/mm using a shimming coil. $\omega_z/2\pi$ of our trap is 50 Hz and $\xi \sqrt{n} < 1 \times 10^{-2}$ for average $n$ at $T$ = 140 $\mu$K.

For a linearly polarized trap beam ($\eta =0$), the difference in the potential well depths, $\Delta U = U(F+1) - U(F)$ is approximately $\bar{U}(\Delta_{\rm HF}/ \Delta)$. This corresponds to 71 Hz for our trap. The linewidth of a Rabi transition depends on thermal distribution of atoms and the Stark-induced broadening in full width at half maximum (FWHM) $\Delta f_S (T)$ is theoretically predicted \cite{Cho JKPS 2007} to be $ \sim 2(k_B T/h)(\Delta U/\bar{U})$, where $k_B$ is the Boltzmann constant. When $T = 140$ $\mu$K, it is 30 Hz. We measure the lineshape of the $|2, -2 \rangle$ to $|1, -1 \rangle$ transition while $\eta$ is changed using a $\pi$ pulse of RF field with duration $\tau_{\rm RF}$ = 200 ms. The results are shown in Fig. 1(a) and (b). When $\eta =0$, FWHM is 32.5 $\pm$ 2.9 Hz and it reaches 4.17 $\pm$ 0.03 Hz at $\eta = 0.413$. The uncertainty-limited value $\Delta f_{\rm UL}$ is 4.0 Hz. The solid line in Fig. 1(b) is $\sqrt{\Delta f_{\rm UL}^2 + (f_0 \Delta \eta)^2}$, where $\Delta \eta$ is the deviation from the magic value and $f_0= 80$ Hz, which is consistent with $\Delta f_S (T)$. Figure 1(c) shows the Rabi lineshape at $\eta = 0.413$ when $\tau_{\rm RF}$ = 2 s; its FWHM is 0.59 $\pm$ 0.02 Hz. Uncorrected broadening corresponds to the error $\delta \eta = 5.4 \times 10^{-3}$ if we disregard other broadening mechanisms such as magnetic field fluctuation. Optimal $\eta$ calculated from the published data \cite{Li matrix elements} is 0.39. However, this value has a large uncertainty because $\beta$ is the difference between the $D1$ and $D2$ couplings of a similar size, and there is a large loss of significant figures.

We measure the coherence time $\tau_c$ of a superposition state using the Ramsey method. A $\pi/2$ pulse of duration $\tau_{\rm RF}$ with detuning $\Delta \omega$ is applied at $t=0$, and another $\pi/2$ pulse is applied at $t=t_d$. Figure 2(a) and (b) show the Ramsey signals when $\eta =0$ ($\tau_{\rm RF}$ = 0.2 ms, $\Delta \omega = 2\pi \times 230$ Hz) and $\eta =0.413$ ($\tau_{\rm RF}$ = 0.5 ms, $\Delta \omega = 2\pi \times 7$ Hz), respectively. When the trap beam is linearly polarized, inhomogeneous broadening dominates and the signal is fitted to $[1+e^{-t_d^2/\tau_c^2}\cos(\Delta \omega t_d)]/2$ with $\tau_c = 17.9 \pm 0.9$ ms \cite{Meschede PRA 2005}. At the magic polarization, homogeneous broadening dominates and the signal follows $[1+e^{-t_d/\tau_c}\cos(\Delta \omega t_d)]/2$ with $\tau_c = 820 \pm 60$ ms. Figure 2(c) shows $\tau_c$ while  $\eta$ is changed. The solid line is from $1/\tau_c = 1/T_2^* +1/T_2^\prime$, where $T_2^*$ and $T_2^\prime$ are from inhomogeneous and homogeneous broadenings, respectively. $T_2^*$ is inversely proportional to $\Delta \eta$, and $T_2^*$ = 6 ms$/\Delta \eta$ and $T_2^\prime$ = 1.5 s fit the data. $\tau_c$ = 820 ms implies that $T_2^*$ = 1.8 s and $\delta\eta$ = $3.3 \times 10^{-3}$. $T_2^\prime$ is mostly limited by temporal fluctuation of the magnetic field $\delta \! B (t)$, which is from unshielded ambient field and noise of the Helmholtz-coil current. In the presence of $\delta \! B (t)$, the Ramsey signal is $S(t_d)=[1+\cos (\Delta \omega t_d - \delta \phi (t_d))]/2$, where $\delta \phi (t) = (\Delta \mu_B/\hbar) \int_{0}^{t} \delta \! B(t^\prime) dt^\prime$ with $\Delta \mu_B = g_F(m_F+m_F^\prime)\mu_B$. When the field fluctuation is random, $\delta \phi (t)$ follows a normal distribution with a standard deviation $(\Delta \mu_B/\hbar) \kappa \sqrt{t}$, where $\kappa$ characterizes the field noise. When averaged over the distribution, $\langle S(t_d) \rangle = [1+e^{-t_d/T_2^\prime}\cos (\Delta \omega t_d)]/2$, where $T_2^\prime = 2\hbar^2/(\kappa \Delta \mu_B)^2$. $T_2^\prime$ = 1.5 s corresponds to field noise $\kappa = 88$ nG/$\sqrt{\rm Hz}$ or current noise of 80 nA/$\sqrt{\rm Hz}$ for our Helmholtz coil. $T_2^\prime$ is inversely proportional to $\Delta \mu_B^2$ and for the $|1, 1 \rangle$ to $|2, 1 \rangle$ transition, we observed $\tau_c$ in excess of 1.5 s.  We did not use the transition for the $\tau_c$ measurement because the Ramsey signal was skewed due to the short lifetime (5 s) of the $|2, 1 \rangle$ state.

A far-detuned optical trap in ultrahigh vacuum can hold atoms for tens of seconds without decoherence from photon scattering. We demonstrate that with magic polarization the trap can be free from Stark-induced broadening and dephasing as well. The magic trap can benefit precision measurement of, for example, the permanent electric dipole moment \cite{Hinds}, where one searches for a small frequency shift. A 1D optical lattice with magic polarization can be easily formed by either a retro-reflection or using a Fabry-Perot cavity \cite{Grimm}. Reflection-induced birefringence is small and it can be corrected \cite{PBC}. The magic lattice is ideal for QIP using neutral atoms. For example, single atoms can be loaded or addressed in a site-specific way by using a site-dependent frequency shift produced by a magnetic field gradient. Previous attempts were plagued by side effects from using a large gradient to overcome the Stark-induced broadening \cite{Meschede NJP}. Sideband cooling to the 3D ground state is difficult because the Lamb-Dicke condition is not satisfied for transverse motion in a 1D lattice. With the extremely narrow line from the magic polarization, it is conceivable to selectively transfer atoms to a specific motional state by using anharmonicity of a Gaussian well in a bottom-up manner. Loading a specific number of atoms to each site is also a challenge, and we may use a collisional shift for a blockade scheme \cite{Rydberg blockade}. For an optical lattice formed by retro-reflecting the Gaussian beam used in our experiment with 1 mK well depth at an antinode, the collisional shift $\Delta f_c$ for a pair of Li atoms in the 3D ground state is 600 Hz \cite{Zoller PRL}. The linewidth we obtained is three orders of magnitude smaller than this. We note that the coherence time we observed is longer than the typical operation time, $1/2\Delta f_c$, of a collisional quantum gate by the same order. The fermionic isotope of lithium, $^6$Li would provide other possibilities with the Pauli exclusion principle and the small hyperfine splitting of 228 MHz.

We thank T. H. Yoon for help with the beam machine and the fiber amplifier, Y. S. Kim for the RF antenna, and S. E. Park for valuable discussions. This work was supported by the National Research Foundation of Korea (Grant No. F01-2009-000-10160-0).

\newpage

{\Large {\bf Figure Captions}}

\begin{itemize}
\item FIG. 1. (Color online) Lineshape of a Rabi transition from the $| 2S_{1/2}, F= 2, m_F=-2 \rangle$ to the $|2S_{1/2}, F= 1, m_F=-1 \rangle$ state. (a) Pulse duration $\tau_{\rm RF}$ is 200 ms. From left to right, the degrees of circularity $\eta$ of the trap beam are 0, 0.292, 0.413, 0.515, and 0.754. The signals are normalized to the peak value at $\eta = 0.413$. The solid lines are Lorentzian functions except for $\eta = 0.413$, where we use the Rabi formula for a square pulse. (b) Full width at half maximum of the lineshapes. The dotted line represents the uncertainty limit. (c) The lineshape at $\eta = 0.413$ with $\tau_{\rm RF} = 2$ s; FWHM = 0.59 $\pm$ 0.02 Hz. The signal is normalized to the peak value in (a). We use a Lorentzian function to fit the data because inhomogeneous broadening is comparable to the uncertainty-limited linewidth.
\item FIG. 2. (Color online) Ramsey signal to measure the coherence time $\tau_c$ of a superposition state. (a) When the trap beam is linearly polarized. $\tau_{\rm RF}$ = 0.2 ms and $\Delta f$ = 230 Hz. Time constant $\tau_c$ of the Gaussian envelope is 17.9 $\pm$ 0.9 ms. (b) For the magic polarization, $\eta = \ 0.413$, with $\tau_{\rm RF}$ = 0.5 ms and $\Delta f$ = 7 Hz. Time constant of the exponential decay is $820 \pm 60$ ms. (c) The coherence time while trap beam polarization is changed. The solid line is from $1/\tau_c = 1/T_2^* + 1/T_2^\prime$, where $T_2^*$ and $T_2^\prime$ are from inhomogeneous and homogenous broadenings, respectively.

\end{itemize}

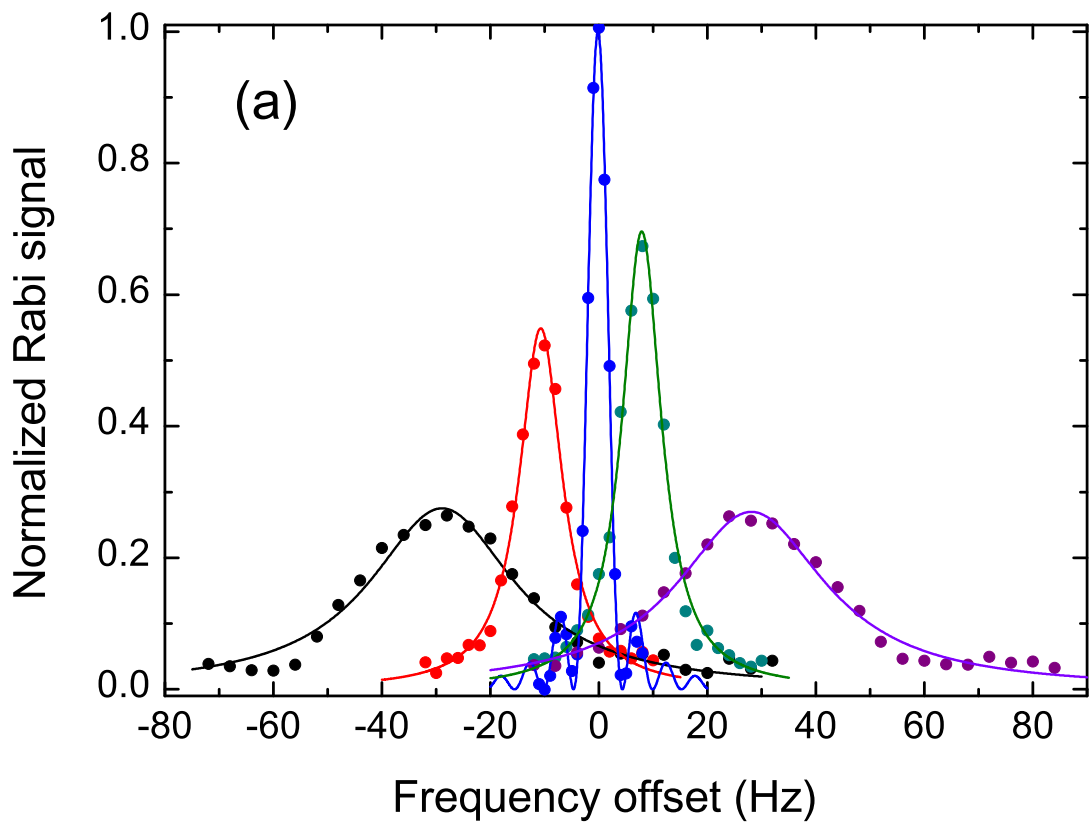
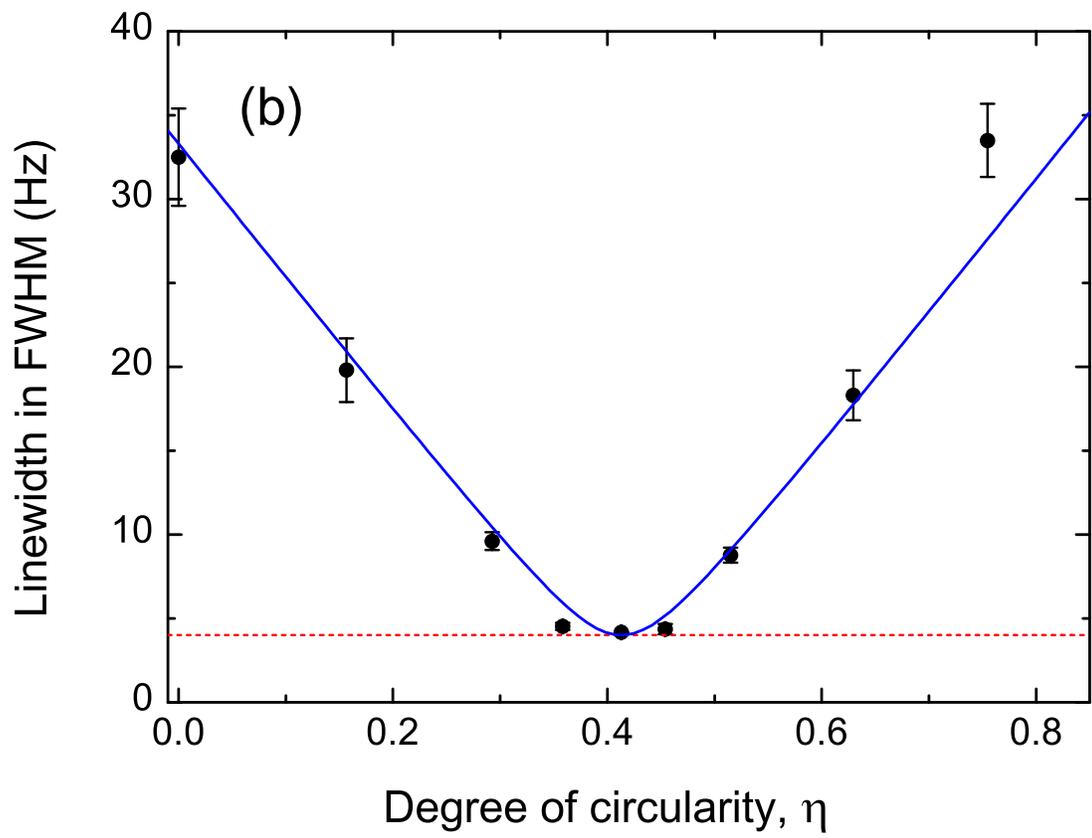
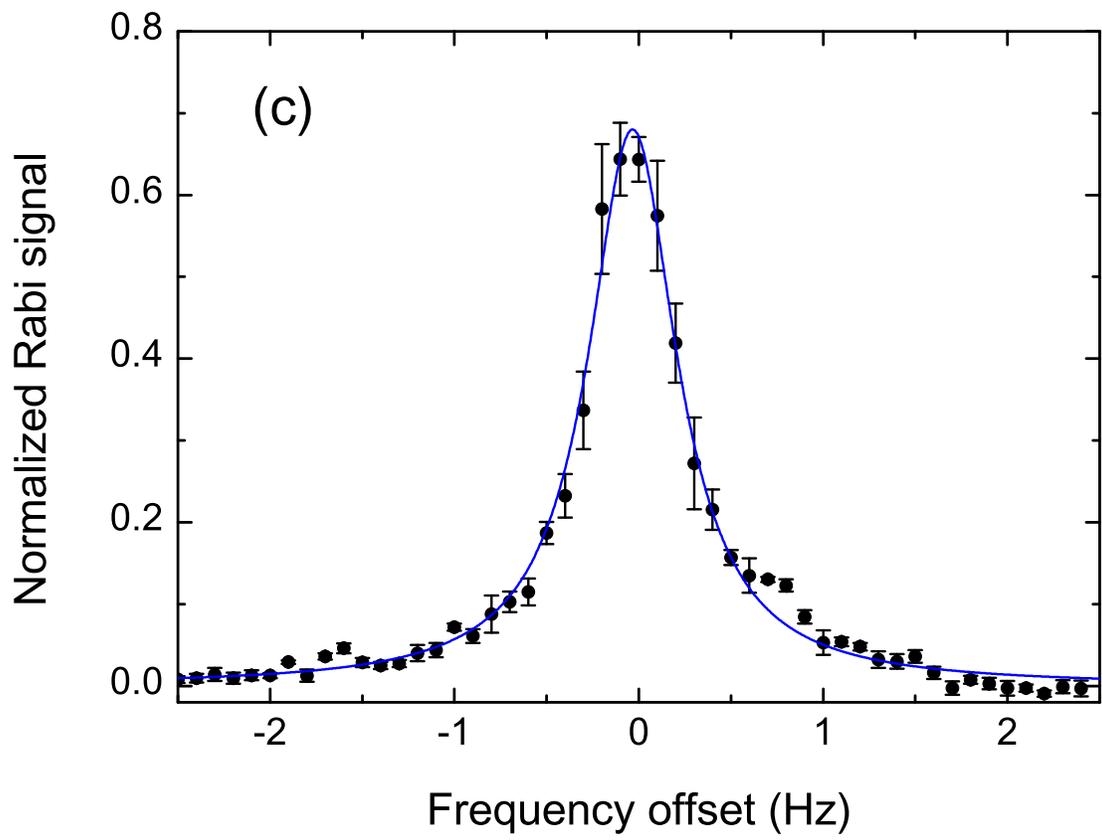
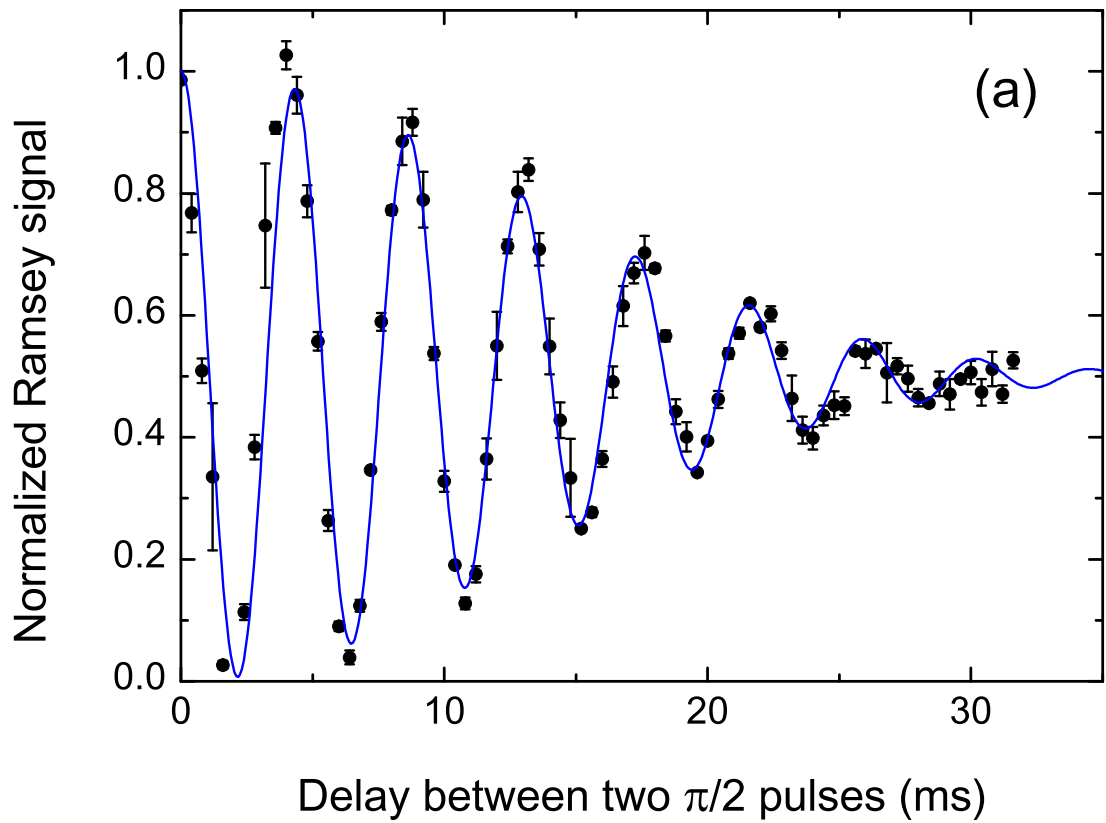
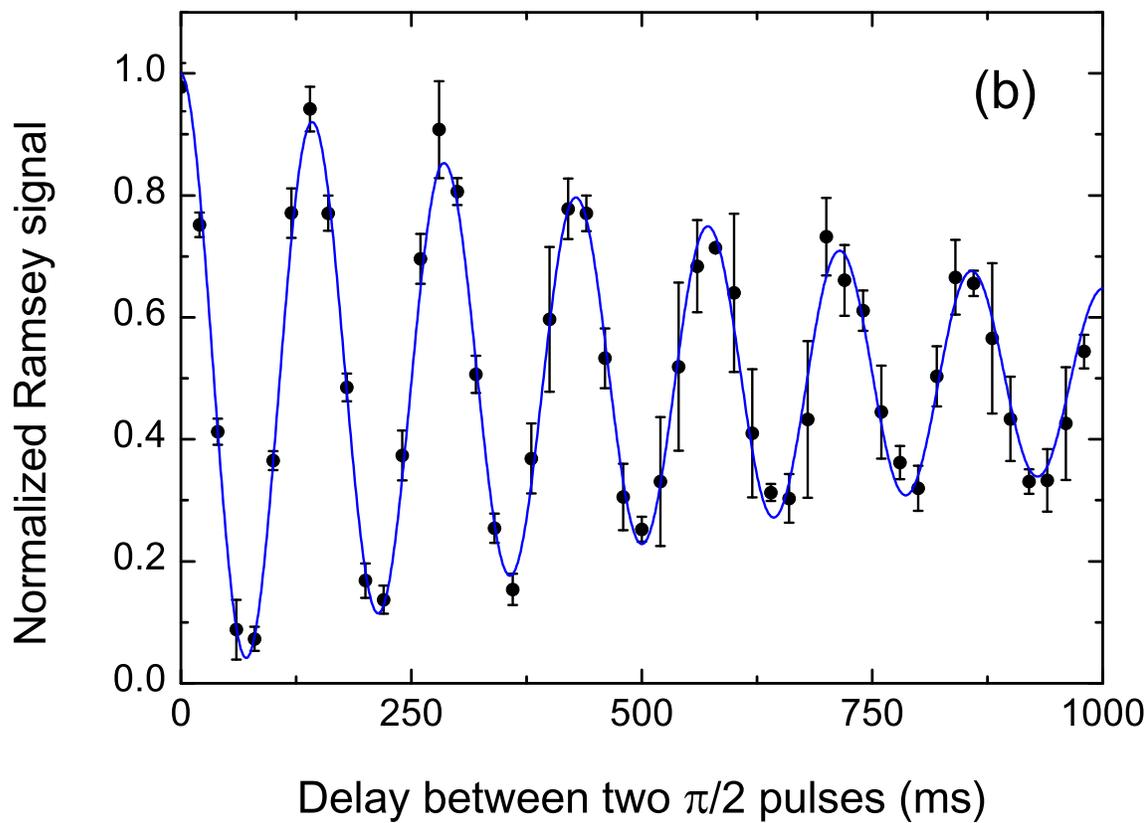
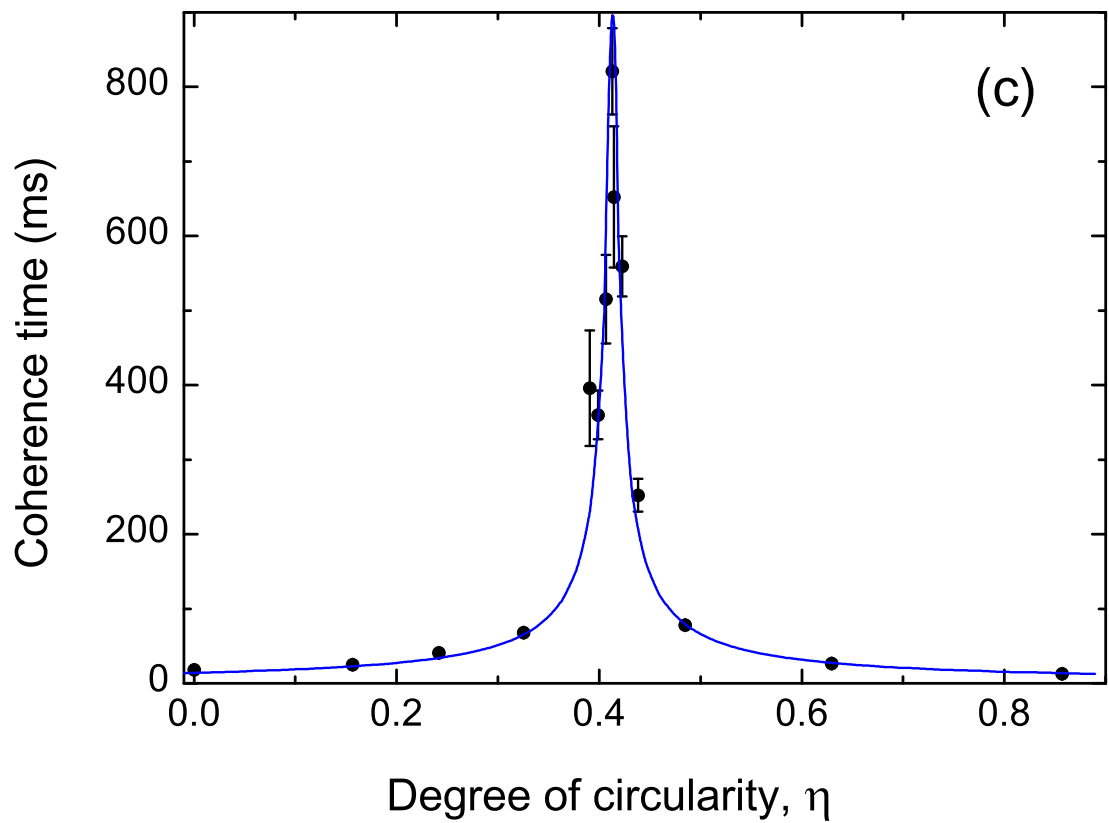
\end{document}